 \documentclass[prd,floats,preprint,superscriptaddress,eqsecnum,floatfix,nofootinbib,12pt,tightenlines]{revtex4}
\pdfoutput=1
\usepackage{amsmath}
\usepackage{graphics, graphicx}
\usepackage{epsfig}
\usepackage{ulem}
\usepackage{color}

\def\cO{{\cal O}}

\def\1p{{(1p)}}
\def\be{\begin{equation}}
\def\ee{\end{equation}}
\def\beq{\begin{eqnarray}}
\def\eeq{\end{eqnarray}}
\def\p0{\phi_0}
\def\z0{\zeta_0}

\def\3G{^3{\cal G}}

\def\ol2{\frac{1}{\ell^2}}

\def\NB{{\rm NB}}
\def\EI{{\rm EI}}
\def\FT{(I,\Psi)}
\def\SP{P(\Dell|\Lambda, Q)}
\def\St0{\Sigma(t_0)}
\def\JJ{(\Lambda, Q)}
\def\Dell{D_{\rm loc}}
\def\Dl0{D_{\rm loc}}
\def\Dobs{D_{\rm uloc}}
\def\Dloct{\tilde D_{\rm loc}}
\def\Dr{D_{\rm r}}
\def\Dobsaoi{D^{\ge 1}_{\rm uloc}}
\def\co{{\cal O}}
\def\bigstar{\mbox{\Large $*$}}
\def\bigdot{\mbox{\Large $\bullet$}}
\def\uf{}
\def\qf{}
\def\zf{}

\def\tf{}

\def\wf{}
\def\nf{}
\def\mf{} 
\def\kf{}
\def\pf{}
\def\tf{}

\newcommand{\ttle}[1]{{\it #1}}

\begin{document}

\vspace{1cm}

\title{Anthropic Bounds on  $\Lambda$\\ from the No-Boundary Quantum State}

\author{James  Hartle}
\affiliation{Department of Physics, University of California, Santa Barbara,  93106, USA}
\author{Thomas Hertog}
\affiliation{Institute for Theoretical Physics, KU Leuven, 3001 Leuven, Belgium}

\bibliographystyle{unsrt}

\vspace{1cm}

\begin{abstract}

We show that anthropic selection emerges inevitably in the general framework for prediction in quantum cosmology. There the predictions of anthropic reasoning depend on the prior implied by the universe's quantum state. To illustrate this we compute the probabilities specified by the no-boundary wave function for our observations at the present time of the values of $\Lambda$ and $Q$ in an inflationary landscape model in which both quantities vary. Within the anthropic range of values the no-boundary state  yields an approximately flat distribution on $\Lambda$ and strongly favors small values of $Q$. This restores Weinberg's successful prediction of $\Lambda$.

\end{abstract}

%\vskip.8in
%\centerline{\Huge  DRAFT -- anthb16 \today} 
%\vspace{1cm}


\maketitle

\tableofcontents

%%11111111111111111111111111111111111111111111111111111111111111111111111111111
\section{Introduction}
\label{intro}

Probabilities predicted by theory for our observations of the universe {\kf must necessarily take into account} a description of the observational situation. That description will include any data that locates the observation in spacetime. But it also includes a description of the observers making the observation i.e. a description of us.  A  simple corollary of this conditioning is that we won't observe what is {\wf inconsistent with our existence. } That is an example of anthropic reasoning. Probabilities for our observations in cosmology are inevitably anthropic.

Anthropic reasoning potentially explains why the observed value of the cosmological constant $\Lambda$ is small when compared to natural values set by the Planck scale as was discussed by {\tf Barrow and Tipler \cite{Barrow86} and by }Weinberg \cite{Wei89}. The universe may be filled with regions where $\Lambda$ is much larger, but we can't live in these. Such large values would not permit galaxies to form. This leads to an {\uf anthropic} upper bound on the value of $\Lambda$ in our region.

Anthropic arguments are meaningful in  a theoretical framework that specifies  both {\uf how observed} parameters can vary as well as prior probability distributions for their values. The predictions of anthropic reasoning {\uf can be}  sensitive to the underlying theory and can thus help to confirm that theory. The example of the anthropic upper bound on the cosmological constant illustrates this: Weinberg \cite{Wei89} showed  { that the anthropic bound on $\Lambda$ agrees well with its observed value in a model in which only $\Lambda$ varies when a {\it uniform} prior probability distribution for it  is assumed over the anthropically allowed range.}  It was pointed out, however, that the agreement with observation becomes worse by several orders of magnitude in theories where the amplitude of the primordial density perturbations $Q$ also varies, assuming a uniform prior probability distribution over both variables \cite{Banks04,Graesser04,LR05,Garriga05,Feldstein05,Tegetal06}. This paper reexamines these predictions from the perspective of quantum cosmology. 

Anthropic selection is often viewed as an application of Weak Anthropic Principle \cite{Barrow86} that is {\uf in addition to a fundamental theory\footnote{See \cite{Schellekens13} for a recent review on anthropic reasoning in particle physics and cosmology.}. Hence it is usually implemented by conditioning on a minimal set of features the universe must possess in order for any sort of life to emerge --- gravitationally bound structures for instance.}

This paper aims first at showing that anthropic {\uf selection} emerges naturally in the general top-down framework for prediction of our observations in quantum cosmology \cite{HH06,Har05} {\it without} the invocation of further principles beyond the basic theory. {\qf Second, we show that, in quantum cosmology, anthropic selection is part of a more general framework for selection. For example, anthropic selection for the cosmological constant is not fundamentally different from the top-down selection for  { classical space-time} and inflation that we have already discussed \cite{HHH10b,H13}. Third, we show by example that the priors supplied by fundamental theories that specify the universe's quantum state can yield predictions that are different and potentially better than the assumption of a uniform prior expressing ignorance of the fundamental theory. 

As an illustration we analyze a successful example of an anthropic prediction in a scalar field theory with a landscape potential that has a large number of minima with different, positive values of the cosmological constant. The minima are surrounded by a range of different inflationary directions that are representative of relatively simple single-field models. The models differ in their predictions for $Q$, the scalar spectral tilt $n_s$ and the tensor to scalar ratio $r$ of the fluctuations, for example.  We model the theory of the quantum state of the universe by the semiclassical no-boundary wave function (NBWF) \cite{HH83}. The NBWF specifies a probability measure on the phase space of classical universes in this landscape that selects inflationary universes \cite{HHH08,HHH08b}. The sum of the probabilities of all universes originating in a given inflationary patch of the landscape implies a relative weighting of the `models' of inflation contained in the landscape, which then yields a prior over various observables \cite{HHH10b,H13}.

%%%%%%%%%%%%%%%jj
\begin{figure}[t]
\includegraphics[width=4.0in]{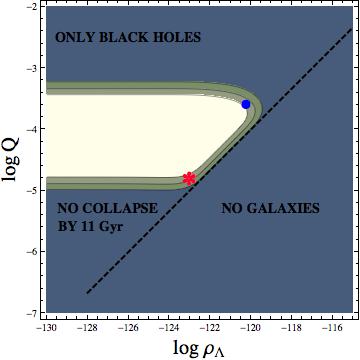} 
\caption{This figure, adapted from \cite{Tegetal06},  shows the factors affecting the selection probability $P(\Dell|\Lambda, Q)$ as quantified by  the fraction $f(\Lambda, Q, t_0)$  of protons in the form of galaxies at a time $t_0=11$Gyr {\kf a few billion years before} the present age, The displayed curves correspond to $f=.4,.5,.6$. The selection probability $P(\Dell|\Lambda, Q)$ is approximately uniform in the white interior region  of the contours shown and decreases quickly outside them. That defines the anthropically allowed range. The most probable values for
$\Lambda$ and $Q$ are indicated by the  {\color{red}\bigstar}\  assuming the no-boundary prior and by the {\color{blue}\bigdot} assuming a uniform one. 
}
\label{fraction1}
 \end{figure}
%%%%%%%%%%%%%%%

All anthropic reasoning concerns probabilities for the observable features of the universe in {\it our} Hubble volume. As a specific example we consider the probabilities for a correlation between three pieces of our data, namely $\Lambda$, $Q$ and data $\Dell$ that describe our nearby universe of galaxies which, in particular, contain data that determine {\pf our present age} from reheating (say).  Correlations are expressed by the joint probability $P(\Lambda, Q, \Dell)$. This may be expressed as 
\be
\label{bayes1}
P(\Lambda, Q, \Dell )=P(\Dell|\Lambda, Q) P_{NB}(\Lambda,Q) .
\ee
This is the form used in traditional anthropic reasoning (TAR)\footnote{ Eq \eqref{bayes1} is the same as Eq (1) in \cite{Tegetal06}.}.

In the traditional terminology the first factor is the `selection probability'. The probability of observing values of $\Lambda$ and $Q$ is negligible when the probability of a galaxy like ours described by $\Dell$  is negligible.  The probability for our local data $\Dell$ is plausibly proportional to the fraction of protons $f(\Lambda,Q, t_0)$ that are in the form galaxies by {\pf the time $t_0=11$Gyr a few billion years before the present age}. Thus we assume
\be
\label{fractgal}
P(\Dell|\Lambda,Q) \propto f(\Lambda, Q, t_0) .
\ee
Figure \ref{fraction1} shows the fraction $f(\Lambda, Q,t_0)$ based on the astrophysical calulations of \cite{Tegetal06} with a sharp cutoff 
that the pregalactic halos form by $t_0 \sim 10^{11} {\rm yrs}$. The white central region defines the anthropically allowed ranges for $\Lambda$ and $Q$. 

The second factor in \eqref{bayes1} is called the `prior' in TAR. The prior in \eqref{bayes1} is determined by the NBWF. Over the narrow anthropic range allowed for $\Lambda$ it turns out to be essentially constant. But it has a significant dependence on $Q$.  We show that directions in our model  landscape that allow for eternal inflation provide the dominant contrbutions to $P_{NB}(\Lambda,Q)$ leading to a $Q$ dependence of the form [cf. \eqref{plamq}]. 
\be
P_{NB}(\Lambda,Q) \propto \exp{({\rm const} /Q^\alpha})
\label{Qdep}
\ee
for a positive constant and power $0< \alpha<2$. 

The NBWF thus strongly favors the smallest value of $Q$ and the largest values of $\log_{10} \Lambda$ in the anthropically allowed region of Figure \ref{fraction1}. This is indicated by the red asterisk ${\color{red}\bigstar}$ . These predictions are in order of magnitude agreement with observations. In effect, the sharp dependence of the NBWF prior has reduced a two-dimensional problem to a one-dimensional one at the smallest allowed value of $Q$ and restored Weinberg's original successful anthropic argument for $\Lambda$ \cite{Wei89}. 

We can contrast this result with the use of a uniform prior over the anthropically allowed region of Figure \ref{fraction1}  such as  that employed in \cite{Tegetal06} to express ignorance of the fundamental theory. The most probable values are indicated by the red asterisk in Figure \ref{fraction1}. These disagree with observation, both for $\Lambda$ and $Q$.

The improvement in predictions for $\Lambda$ and $Q$ that follow from using the NBWF prior is one of the important results of this paper. But it is not the only one. Another result is how anthropic reasoning is implemented within quantum cosmology in terms of probabilities for observations conditioned on a description of the observational situation. 

A theory in quantum cosmology is specified by a theory of the quantum state of the universe ($\Psi$ for short) and a theory of its dynamics ($I$ for short). These predict the probabilities for members of the ensemble of entire possible classical histories of the whole universe \cite{HHH08,HHH08b}. The kind of question we have to address is how do we get from these probabilities for histories to the probabilities for observations in our Hubble volume that enter into \eqref{bayes1}?  Indeed, what do we mean by `our' Hubble volume in a theory that permits classical histories with large or even infinite volumes in which our data will  be replicated over and over again in different places?  How do we treat ourselves as quantum physical systems within the universe with only a probability to exist in any one Hubble volume?  How do we derive \eqref{Qdep} from $\FT$? 
It is this kind of question with  which most of the rest of the paper is concerned.

This paper is organized as follows: Section \ref{small} illustrates how anthropic constraints arise automatically in quantum cosmology. A simple, if unrealistic, model landscape is used in which matter driven eternal inflation is prohibited and, as a consequence, all universes are `small'. The analysis is close to traditional anthropic reasoning. The rest of the paper is mostly devoted to {\kf anthropic reasoning in} landscape models with regions of eternal inflation which give rise to very large universes. Section \ref{landscape} describes the landscape model assumed. Section \ref{measure} reviews how the NBWF predicts the probabilities of an ensemble of classical histories.  Section \ref{typicality} reviews the typicality assumptions that are necessary to describe `us' in very large universes where, as physical systems, we may be replicated in many different places. The important connection is made between first person probabilities for our observations and top-down probabilities conditioned  on at least one instance of part of our data. The calculation of these top-down probabilities is developed in Section \ref{pred}. Eternally inflating histories are selected by these probabilities because there are more places for our data to occur in them than in small universes. Histories starting at the lowest exit from eternal inflation are shown to dominate all others. These results considerably simplify the implementation of anthropic selection which is done in Section \ref{ourHV}.  The results are given in Section \ref{together} where we show that Weinberg's \cite{Wei89} bounds on $\Lambda$ are restored by the NBWF even though $Q$ is allowed to scan. Section \ref{testing} discusses  the process of anthropic reasoning in quantum cosmology more generally and its role  in testing physical theories. Section \ref{summary} summarizes the argument as a series of selections of which the anthropic one is but one step.

%%22222222222222222222222222222222222222222222222222222222222222222222222222222222222222222

\begin{table}[t]
\label{ourdata}
\begin{tabular}[t]{|l|p{5.5in}|}
\hline 
%\multicolumn{2}{|c|}{}\\
\multicolumn{2}{|c|}{ \large \bf Our Data  }\\ \hline
%\multicolumn{2}{|c|}{}\\ \hline
$D$  &all of our data about the universe on all scales   \\ \hline  
%$\Dr$ & kk data which is  $D$ minus records of observations that we aim to compute probabilities for. \\ \hline
 $\Dobs$ &  ultralocal data on planet size scales including a description of us  as physical systems observing the universe and excluding any records of observations of quantities whose probability we aim to compute e.g.  $\Lambda$ and $Q$. \\ \hline
$\Dell$ & local data on the scales of our galaxy and others nearby including those which together with $\Lambda$ determine a FLRW model and, in particular, the present age.  \\ \hline
$\Dr$& A convenient notation for the union of $\Dobs$ and $\Dell$. \\ \hline
\end{tabular}
\caption{Different pieces of our data used throughout the paper.}
\end{table}

\section{Anthropic Selection in Small Universes}
\label{small}

This section uses a simplified  model  to illustrate how anthropic selection emerges automatically  in quantum cosmology from the probabilities for our observations that are  predicted by a quantum state of the universe. In particular we illustrate how the probabilities relevant for selection are connected to the probabilities for {\kf classical} histories of the universe arising from $\FT$. 

We work in a minisuperspace of homogeneous, isotropic, spatially closed geometries. For the matter we assume a multicomponent scalar field $\vec \phi$ moving in a landscape described by a potential $V(\vec\phi)$. The potential has different minima with different values of $\Lambda$. These are approached from  different directions $J$ that are each approximately one dimensional and separated by steep barriers. Each direction is therefore equivalent to a one dimensional scalar field $\phi^J$ moving in a potential $V^J(\phi^J)$ with a minimum characterized by a cosmological constant $\Lambda=V^J(0)$. Landscapes of this kind are discussed in more detail and specificity in Section \ref{landscape}.  The potentials for different directions $J$ will also differ in other parameters besides $\Lambda$. To keep the discussion manageable we will assume that there is only one further parameter that we call  $\mu$. The landscape provides a mechanism for $\Lambda$ and the other parameter to vary (scan).

Most importantly, for this model we will assume that none of the potentials $V^J$ allows for scalar field driven eternal inflation. This simplifies the discussion because it means that the {constant density} spatial slices of the predicted histories will be manageably finite. { Thus the ensemble consists of ``small universes" only}. Such landscapes may have trouble predicting an adequate number of inflationary e-folds for realistic structure formation, but they provide a simple model to illustrate anthropic reasoning. {\kf We consider large, eternally inflating universes later in the paper.}

As will be described in Section \ref{measure}, the NBWF in its semiclassical approximation predicts probabilities for the members of an ensemble of classical histories --- histories correlated in time by classical dynamical laws. Individual histories are characterized by the direction $J$  they roll down from and by the approximate value of the scalar field   $\p0^J$ from which they start to roll.  A particular direction will predict a specific value for the  amplitude $Q(J,\p0^J)$ which can be found from an analysis of the linear fluctuations away from homogeneity and isotropy. We can use $Q$ instead of  $\mu$ for specifying $J$ . Thus classical histories can be labeled by $(\Lambda, Q, \p0)$ (replacing $J$ by the more specific $(\Lambda, Q)$). The NBWF supplies probabilities  $P_{\NB}(\Lambda, Q, \phi_0)$ for these four-dimensional histories of spacetime  geometry and matter fields. 

Probabilities for observation are deduced from probabilities of histories by conditioning on a description of the observational situation described by our data, taking in account its possible locations \cite{HH09}. We denote by $D$ all of the data we have about the universe, on all scales, at any one time: every record of every experiment, every astronomical observation of distant galaxies, every available description of  every leaf, etc., and necessarily every piece of information about ourselves. The choices of coarse graining used to define this information, the level of accessibility demanded, etc., are inevitably subjective but assumed here to be fixed.  

Our data $D$ can be divided into a range of scales.  On the scales of our planet we have a large amount of very detailed data $\Dobs$ that (at a coarse-grained level) describe the telescopes and satellites that make cosmologically relevant observations. But they also describe the surface features of the Earth, its biota, and us, both individually and collectively as physical systems within the universe and a great many other things. These data also include the records of our observations of $\Lambda$ and $Q$ but we leave these out of $\Dobs$ because we aim at predicting probabilities for these quantities as if they were not yet observed.

In addition to data on planet size scales we also have data $\Dell$ describing the nearby universe in our Hubble volume\footnote{We try to adhere to the convention that ``global'' means the whole universe including the parts outside our Hubble volume, ``local'' means within our Hubble volume, and ``ultralocal'' means on scales of the planet.}. This includes information about stars, our galaxy,  and galaxies other than our own. In particular $\Dell$ contains data that together with $\Lambda$ would specify a FLRW model. These include the Hubble constant, the matter density, the temperature of the CMB, etc.  The data $\Dell$ and $\Lambda$  determine our time $t_0(\Dell,\Lambda)$ as measured, say, from reheating. 

These three kinds of  data will be used throughout the paper not just in this section\footnote{Some readers may find it helpful to think of the  various levels of data as what was available at various times in the history of physics, 1850, 1900, 1945, etc. However this is open to contention on exactly what was available at these times.}.
They are summarized in the Table. 
   
The quantum mechanical probability $P_g(\Dobs)$ that $\Dobs$ occurs in one galaxy is an extraordinarily small number. It is also well beyond the power of present day physics to compute not least because it involves the probabilities of the chance events of biological evolution. {On the plus side, $P_g(\Dobs)$ is plausibly  independent of $\Lambda$ and $Q$ --- the observables we aim to predict.} We will assume in this Section that the histories in the classical ensemble represent universes that are small enough that the probability that there is more than one instance of $\Dobs$ is negligible. We are unique physical systems in any history in this model {\mf if we exist at all}.

We investigate the predictions of the NBWF for anthropic correlations among the observables ${\cal O} \equiv (\Lambda, Q, \Dell)$ conditioned on the requirement that the universe contains $\Dobs$ {\mf in particular that it contains us}. That is, we calculate $P(\Lambda, Q, \Dell|\Dobs)$. 
 From the definition of conditional probability and using $\Dr\equiv(\Dobs,\Dell)$ we have:}
\begin{align} 
\label{histsum}
 P(\Lambda, Q, \Dell |\Dobs) &\propto P(\Dr|\Lambda, Q) P(\Lambda, Q)    \nonumber \\
 & \propto \int d\phi_0 P(\Dr |\Lambda, Q, \p0) P(\Lambda, Q, \p0) . 
\end{align}
Thus the probability for correlation is expressed in terms of the NBWF probabilities for histories in the landsape. 

Since $\Lambda$ is a condition in \eqref{histsum},  we know the age and the time of galaxy formation  $t_0(\Dell,\Lambda)$.  
The time $t_0$  specifies a  spacelike surface $\Sigma(t_0)$ in each history contributing to \eqref{histsum}. Our Hubble volume is somewhere on this surface. 
The probability for $\Dr$ to exist in some Hubble volume on the surface $\Sigma(t_0)$ is the probability $P_E$ for it to exist in any one Hubble volume multiplied by the number of Hubble volumes on $\Sigma(t_0)$ which we denote\footnote{If $\Sigma(t_0)$ does not occur in a history we take $N_h=0$.}  
 by $N_h(\Dell,\p0, \Lambda, Q)$. While the probability $P_E$ for $\Dr$ to occur in any Hubble volume clearly depends on $\Lambda$ and $Q$ it is plausibly independent of $\p0$ --- the start of slow roll inflation --{\kf at least for $\p0$ sufficiently large}. Assuming this we find
\be
\label{globtolocal}
P(D_r|\Lambda,Q,\p0) \propto N_h(\Dell,\p0,\Lambda,Q) P_E(\Dr|\Lambda, Q) .
\ee
Inserting this in \eqref{histsum} we find 
\be
\label{tar}
P(\Lambda, Q, \Dell |\Dobs) \propto   P_E(\Dr|\Lambda,Q )P_{\NB}^{\rm VW}(\Lambda,Q) .
\ee
Here,  $P^{\rm VW}_{\NB} (\Lambda, Q)$ is the volume weighted NBWF prior. Larger volumes are favored because there are more places for $D_r$ to be \cite{Page97,HHH08b,HH09}.  

We are now in the same position as with \eqref{bayes1} in the Introduction. The probability $P_E(\Dr|\Lambda,Q )$ acts as the selection probability and $P_{\NB}^{\rm VW}(\Lambda,Q)$ is the prior. Of course \eqref{Qdep} doesn't hold because the model doesn't allow for eternal inflation.  

This simple model offers a clear illustration how TAR is part of the general framework for prediction in quantum cosmology, and how both selection and prior probabilities  arise from the fundamental theory $\FT$. But the  assumption that all universes in the ensemble are ``small" is too restrictive for realistic application. Even simple potentials like $V=(1/2)m^2 \phi^2$ have a regime of eternal inflation. And, as we will see, the resulting large universes will dominate the probabilities for anthropic selection.

For large universes, the approximation $N_h P_E \ll 1$ that we are unique breaks down. The data $D$ may be replicated many times in a large universe in different Hubble volumes. Different techniques and stronger assumptions are then required {to implement anthropic selection}. The next sections are devoted to these.

%%33333333333333333333333333333333333333333333333333333333333333333333333333333333333333333333
\section{A Landscape Model}
\label{landscape}

Motivated both by observation \cite{Perlmutter99,Planck13} and fundamental theory \cite{Bousso00,Denef04} we consider a landscape model consisting of an ensemble of minima with different, positive values of the cosmological constant $\Lambda$. We assume each minimum is surrounded by the same, multi-dimensional scalar potential exhibiting a range of different potential behaviors in different directions in field space. This includes directions where the slow-roll conditions for inflation are satisfied for a range of field values. We assume that different inflationary directions are separated by steep barriers so that each direction can be described in terms of an effective single-field model. 

Our landscape contains a number of families of inflationary potentials that are representative of single-field potentials with only a few free parameters. One class of directions comprises power law potentials of the form 
\be
\label{pwlaw}
V(\phi) = \Lambda+\lambda \phi^n, \qquad n >0 . 
\ee
This constitutes a two parameter family of inflationary potentials and a three parameter family of directions labeled by $J=(\Lambda, \lambda, n)$. 

We also consider directions with plateau-like potentials of the form
\be
\label{plat}
V(\phi) = \Lambda + V_0 (1-\phi^p/\mu)^2, \quad p \geq 2
\ee
which constitute a four parameter family of directions labeled by $J=(\Lambda, V_0, \mu, p)$.  Finally we include a weighting factor $g(J)$ that accounts for the relative frequency with which different potentials occur in the landscape. We will assume however that our landscape has no exponentially strong statistical bias that (dis)favors a particular inflationary potential.

In this toy landscape we find ourselves near one of the minima of the potential at the present time. If the field was displaced from the minimum in the past then our universe may have undergone a period of scalar field driven inflation. Our observations will then depend on which direction in field space the early inflationary universe rolled down from. The structure of the landscape alone is not sufficient to predict what we should see. To proceed we must first complete the theory with a model of the quantum state of the universe.

 %%444444444444444444444444444444444444444444444444444444444444444444444444444444444444444444
\section{No-Boundary Measure}
\label{measure}
A quantum state of the universe is specified by a wave function $\Psi$ on the superspace of  geometries ($h_{ij}(x)$)  and matter field configurations ($\phi(x)$) on a closed spacelike { three}-surface $\Sigma$. Schematically we write $\Psi=\Psi[h_{ij},\phi]$. 
A wave function of the universe $\Psi[h_{ij},\phi]$ does not predict a single universe. Rather it predicts an ensemble of possible classical histories of the universe along with their quantum probabilities.

We adopt the no-boundary wave function (NBWF) \cite{HH83} as a model of this state. In the semiclassical approximation the NBWF is given by\footnote{We use Planck units where $\hbar=c=8\pi G=1$.} \cite{HHH08}\footnote{\kf The NBWF was originally defined \cite{HH83} as a path integral over a complex contour of four-geometries and matter field configurations \cite{HH90}. Its semiclassical approximation is then given by the saddle points that contribute to the steepest descents distortion of this contour. As shown in \cite{HL90} different contours could in principle give different semiclassical approximations. Here we restrict to the wide class of NBWF contours in which the dominant saddle point of the action supplies the leading semiclassical approximation. That is consistent, indeed mandated, by the holographic definition of the semiclassical NBWF \cite{HM04,HH11}.}
\begin{equation}
\Psi[h_{ij},\phi] \approx  \exp(-I[h_{ij},\phi]) = \exp(-I_R[h_{ij},\phi] +i S[h_{ij},\phi])
\label{semiclass}
\end{equation}
where $I_R[h_{ij},\phi]$ and $-S[h_{ij},\phi]$ are the real and imaginary parts of the Euclidean action $I$ of a compact {\it regular} saddle point solution that matches the real boundary data $(h_{ij},\phi)$ on its only boundary $\Sigma$. If $S$ varies rapidly compared to $I_R$  the wave function takes a WKB form and predicts the boundary configuration $(h_{ij},\phi)$ evolve as a Lorentzian, classical universe. The ensemble of classical universes predicted by the NBWF is of particular interested in cosmology, since classical spacetime is a prerequisite for cosmology as we know it. Our existence {\it selects} the quasi classical realm of the wave function of the universe\footnote{\kf The NBWF in principle predicts the probabilities of other sets of alternative coarse-grained histories --- other realms --- besides the quasiclassical ones that are the exclusive focus of this paper.  (See e.g. \cite{GH07}.) This focus is appropriate for two reasons. First, we as physical systems within the universe are part of its quasiclassical realm --- described in terms of quasiclassical variables. Second, the variables of concern --- the cosmological constant, the CMB fluctuations, and the age --- are all measured by the classical behavior of the universe. One could imagine anthropic probabilities for highly non-classical variables but we are not concerned with these in this paper.}.

In our model landscape the NBWF predicts an ensemble of classical Friedman-Lema\^itre-Robertson-Walker (FLRW) backgrounds with Gaussian perturbations \cite{HHH10a}. The NBWF has the striking property that it predicts only FLRW backgrounds which undergo some amount of matter driven, slow-roll inflation \cite{HHH08,HHH08b}. Intuitively this is because only universes with sufficiently small gradients at early times can be smoothly glued onto regular saddle points. Hence the NBWF populates the landscape in a very specific manner: it {\it selects} the landscape patches where the conditions for inflation hold \cite{HHH10b}. 

The model landscape we consider has only single-field inflationary directions. In each of these the NBWF selects a one-parameter set of inflationary backgrounds, which can be labeled by the direction $J$ of the landscape it rolls down from and by the absolute value $\phi_0^J$ of the scalar field  at the `South Pole'  (SP) of the associated saddle point (where $a \rightarrow 0$) \cite{HHH08}. It turns out that $\phi_0^J$ is approximately equal to the largest value of the scalar field in the corresponding real, classical background. Hence the number of efolds $N(\phi_0^J) \approx \int_{\phi_e^J}^{\phi_0^J} (V/V')$ where $\phi_e^J$ is the value at which $\epsilon^J=1$ and hence inflation ends. For sufficiently small $\Lambda$ the range of $\phi_0^J$ also has a lower bound at a critical value $\phi_{0c}^J$, with $N(\phi_{0c}^J) \sim {\cal O}(1)$ \cite{HHH08}.

The individual histories in the ensemble are given by the integral curves of $S$ and have tree-level probabilities proportional to $\exp[-2 I_R(h_{ij},\phi)]$. It turns out the probabilities $P_{NB}(\phi_0^J)$ of the backgrounds are proportional to \cite{HHH08}
\be
\exp[-2 I_R] \approx \exp \left[\frac{3\pi}{\Lambda+ V(\phi_0^J)}\right] .
\label{bu}
\ee
The probabilities for scalar curvature perturbations $\zeta^J$ and tensor perturbations $t_{ij}^J$ on $\Sigma$ are specified by the no-boundary wave function of perturbations around homogeneous saddle points \cite{Halliwell85}. In the semiclassical approximation they are given by the usual product\footnote{The regularity condition on the saddle points implies that perturbations start out approximately in the Bunch-Davies ground state \cite{Halliwell85}.} of Gaussian probabilities $P(\zeta_k^J|\phi_0^J)$ and $P(t_{ij(k)}^J|\phi_0^J)$ for fluctuation modes $\zeta_k^J$ and $t_{ij(k)}^J$ on $S^3$ \cite{Halliwell85}. In this paper we are interested in observables associated with scalar perturbations only. We therefore coarse grain over the tensor perturbations.
The NBWF probabilities for $\zeta^J$ imply the usual scalar power spectrum $\Delta^2_{\zeta^J} \equiv (k^3/2\pi^2) {\cal P}^J_{\zeta}(k) = (Q^J)^2 (k/k_{*})^{n^J_s-1}$ where $(Q^J)^2 \sim (V/\epsilon)^J$ and $n^J_s-1 = 2 \eta^J - 6\epsilon^J$, with $\eta^J \equiv (V''/V)^J$, are evaluated at the background value $\phi_{*}$ when the reference scale $k_{*}$ crosses the horizon during inflation. 

In our landscape model, saddle points with $V < \epsilon$ at the SP give rise to an ensemble of nearly homogeneous inflationary universes with small fluctuations. By contrast, saddle points starting in eternal inflation regions of the landscape where $V > \epsilon$ predict high amplitudes for significantly inhomogeneous universes that have large (very) long-wavelength perturbations on the scales associated with eternal inflation. This is because in eternal inflation, the typical size $\Delta_{\zeta^J}$ of perturbations on the horizon scale is comparable or larger than the background field motion in a Hubble time. As a consequence the perturbations have a large effect on the structure of the universe on those scales.

%%%%%55555555555555555555555555555555555555555555555555555555555555555555555
\section{Typicality}
\label{typicality}

This section begins the discussion of anthropic selection in the NBWF when the landscape implies eternally inflating geometries in the ensemble of possible classical histories of the universe. As discussed in Section \ref{measure}, in an eternally inflating history, the geometry of constant density surfaces can become large (or even infinite) and significantly inhomogeneous. We are therefore no longer in the small universe, nearly homogeneous, case of Section \ref{small}. We begin by discussing what changes from the small universe case and what remains the same. 

The data $\Dr=(\Dobs,\Dell)$ of course are exactly the same. The goal of making anthropic selections based on NBWF probabilities for correlations among the observables $(\Lambda, Q, \Dell)$ is unchanged. What has to change is the nature of the conditioning on $\Dobs$ describing the main part of the observational situation because it can no longer be assumed to be rare in eternally inflating histories. 

When we discussed small universes in Section \ref{small} we assumed that there was no more than one instance of $\Dobs$ in any history. In particular we assumed that $N_h P_E \ll 1$. But this inequality will not be satisfied in the very large $N_h$ universes that arise from eternal inflation. Rather, there will be a significant probability that the data $\Dobs$ will occur (be replicated) in many different locations in each history. Indeed, in an infinite universe the probability is unity for any finite number of instances of $\Dobs$. 

As human observers on the planet Earth\footnote{A little more concretely the system could be taken to be the human scientific IGUS, that is the information gathering and utilizing system \cite{Gel94} constituted by the human beings on the planet Earth engaged in the scientific enterprise.} we are one instance of $\Dobs$ in one history. 
Words such as `we', `our', and `us' refer to this particular physical subsystem.  Probabilistic predictions for correlations in our data, for the outcomes of our future observations based on it, etc are called `first-person probabilities'.  As already mentioned in the Introduction, probabilities for anthropic reasoning are first-person. How are {\it our} observations restricted by the requirement that {\it we} can exist?

The theory $T=\FT$ does not predict first-person probabilities directly. Rather it predicts the probabilities for the members of an ensemble of four-dimensional classical histories of the universe. These are called third-person probabilities\footnote{Some readers may be helped by thinking about third-person probabilities as ones that could be verified by a hypothetical observer somehow outside the universe and outside of time  that could observe an entire classical history. This risks confusion because there are no such observers and the quantum mechanics of closed systems does not require them.}.

An individual history may contain the data $\Dobs$ at one or more locations. After all, $\Dobs$ are just a very special kind of fluctuation with only a probability to occur in any one Hubble volume, and, in a large universe, a significant probability to occur in many. We are one of these instances, but the theory $T$ does not specify which one. A further assumption is therefore needed to connect the third-person probabilities supplied by $T$ with the first-person probabilities needed e.g. for anthropic reasoning. {Its specification should be regarded as part of the theoretical framework for prediction \cite{HS09}, testable like any other\footnote{This assumption should not be confused with the measure which is supplied by the quantum state.}.} 
The simplest assumption, and the one we shall make here, is that we are not Boltzmann brains and that we are equally likely to be any of the other incidences of $\Dobs$ that occur in a history. Put differently we assume that what we observe is typical of what is observed by the other incidences. 

All we know is that our universe exhibits at least one instance of $\Dobs$ ($\Dobsaoi$ for short) --- our instance. The first-person probability $P^{\1p}$ for observables $\cO$  (which are $(\Lambda, Q, \Dell)$ in our case) is thus given by the sum over the third-person probabilities for histories with these observables conditioned on the requirement that each history  contain at least one instance of $\Dobs$. 
\be
P^\1p(\cO) = P(\cO|\Dobs^{\ge 1}) .
\label{firstperson}
\ee
The right hand side is a familiar top-down probability \cite{HH06} for $\cO$ which can be calculated once a relation between `us'  and instances of $\Dobs$ has been assumed\footnote{Although $\Dobs$ is the major part of  $D_r=(\Dobs,\Dl0)$ the reader might wonder why we don't condition on $\Dr^{\ge 1}$ which is a slightly less informative typicality assumption as we could do in the small universe case. The answer is that in an infinite universe the probability that there is at least one instance of any part of our data is unity. Conditioning on that data therefore does not select among different infinite universes that emerged from  directions with different values of $\Lambda$ and $Q$.  To have selection for $\Lambda$ and $Q$ it is best to use that part of our data that is independent of them {\nf to specify the observational situation}. That's $\Dobs$. That preserves the part of $D_r$ that depends on $\Lambda$ and $Q$ for use in selection. This kind of choice is {\nf common in science} and discussed more generally in Section \ref{testing}. }.

%%%%%%%%%%%%6666666666666666666666666666666666666666666666666666666666666666666666
\section{Top-Down Probabilities}
\label{pred}

{\nf Top-down probabilities are conditional probabilities relevant for the prediction of our observations \cite{HH06}. Eq. \eqref{firstperson} is an example. 
The top-down requirement in \eqref{firstperson} that at least one instance of $\Dobs$ exist is a global condition. It {\it selects} from among the ensemble of possible classical histories ones in which $\Dobs$ occurs at least once, somewhere, sometime. In this section we show that the top-down selection favors the eternally inflating histories in the ensemble where the universe becomes spatially very large. We use this result to simplify the calculation of the resulting probability $p^\1p(\cO)$  for the observables $\cO =(\Lambda, Q, \Dl0)$ that will be used for anthropic selection in the next section. }

The starting point to evaluate top-down probabilities are the `bottom-up' probabilities \eqref{bu} of the individual universes in the ensemble.
In our model landscape the general form of top-down probabilities $P({\cal O}|\Dobsaoi )$ is given by
\be
\label{Pobs}
P({\cal O}|\Dobsaoi) \propto \sum_J g(J) \int P({\cal O}|\phi_0^J,\zeta_{k}^J) P(\Dobsaoi |{\cal O},\phi_0^J, \zeta_{k}^J) P_{NB}(\phi_0^J, \zeta_{k}^J)
\ee
where the integral is over $\phi_0^J$ and $\zeta_{k}^J$ labeling the histories in direction $J$. The last factor is the NBWF prior given by \eqref{bu} together with the NBWF probabilities $P(\zeta_k^J|\phi_0^J)$ for scalar perturbations. {\qf The first factor can also be calculated from the NBWF by expressing ${\cal O}$ in terms of the variables $\phi_0^J$ and $\zeta^J$ characterizing the histories.} {\zf This factor leads to anthropic selection when ${\cal O}$ describes correlations involving observables referring to us and will be discussed in Section \ref{ourHV} below.}

{\zf The second factor in \eqref{Pobs} is the `top-down' (TD) weighting}. The TD requirement that at least one instance of $\Dobs$ exists somewhere is trivial in histories in the ensemble that are sufficiently large. {\wf After all, in an infinite universe anything occurs somewhere sometime with probability 1.}  In universes that are small however the probability  $P(\Dobsaoi|{\cal O},\phi_0^J, \zeta_{k}^J)$  that $\Dobs$ exists is exceedingly small for realistic $\Dobs$ as discussed in Section \ref{small}. We have argued \cite{HH09} this is the case in saddle point histories that start below the threshold of eternal inflation, which predict high amplitudes for nearly homogeneous universes only, with $N_h \propto e^{3N}$. {\nf In those universes the conditioning on $\cO=(\Lambda, Q, \Dl0)$ in the TD factor means that $\Dobs$ occurs on a spacelike surface $\Sigma$ of approximately constant time $t_0(\Dl0,\Lambda)$. The data $\Dobs$  occur only with a small probability $P_E$ in any of the $N_h(\p0^J,\zeta^J)$  Hubble volumes on this  surface $\Sigma$. Hence in universes with $N_h \ll 1/P_E(\Dobs)$  the TD factor $P(\Dobsaoi|{\cal O},\phi_0^J, \zeta_{k}^J) \approx P_E N_h \ll 1$.}

By contrast, saddle points with a regime of eternal inflation predict high amplitudes for configurations that have large perturbations on scales associated with eternal inflation \cite{HHH10b}. As a consequence the volume of {\nf constant density surfaces in} such universes is generally exceedingly large or even infinite so that $P(\Dobsaoi|{\cal O},\phi_0^J, \zeta_{k}^J) \approx 1$. {\wf We will assume, however, that even though eternally inflating histories are globally inhomogeneous, the local properties are statistically the same in every Hubble volume in each individual history. That means that the observed values of $\JJ$ are the same in each Hubble volume, and that the statistical properties of the fluctuations on sub-Hubble-volume scales are the same in each Hubble volume\footnote{\nf {We are thus ignoring inhomogeneous configurations consisting of quasi-homogeneous patches associated with different directions, since one expects the contribution to the NBWF probabilities of such configurations to be very small.} This means that the typicality assumption discussed in the preceding section is moot, but we aim at a general discussion.}. In particular, the probability that $\Dobs$ occurs in a Hubble volume is the same in all of them. This leads to the following accurate approximation of the TD weighting in directions with a regime of eternal inflation,
\be
P(\Dobsaoi| {\cal O},\phi_0^J, \zeta_{k}^J) \approx  \frac{1}{2}\left[1+\tanh [ \frac{3}{2} (N(\phi_0^J) - N(\phi_{EI}^J)]\right] \equiv P_{TD}(\phi_{EI}^J,\phi_0^J)
\label{td}
\ee
where $\phi_{EI}^J$ is the value of $\phi_0^J$ at the eternal inflation threshold $V^J =\epsilon^J$ in direction $J$. 
This is independent {\zf of $P_E$ and of the precise nature of $D_{\rm uloc}$} and hence computable \cite{HH09}.

{\qf The observables ${\cal O}$ whose probabilities we compute in this paper only refer to features of backgrounds or to the expected values of perturbations around different backgrounds. For example, minima are characterized by different values of $\Lambda$ and the value we observe is determined by which minimum our background winds up at. Similarly the scalar amplitude $Q^2 \sim V/\epsilon$ on the associated scale is specified by the shape of $V$ around the field value probed by the background when the associated scale crosses the horizon during inflation. This means the probabilities for ${\cal O}$ do not explicitly depend on the perturbation variables $\zeta^J$. Hence we can coarse grain (sum) over the probabilities of the perturbations in \eqref{Pobs}.} To leading order in $\hbar$ this yields
\be
\label{Pobs2}
P({\cal O}|\Dobsaoi) \propto \sum_J g(J) \int P({\cal O}|\phi_0^J)\left(1+\tanh [ \frac{3}{2} (N(\phi_0^J) - N(\phi_{EI}^J)]\right)P_{NB}(\phi_0^J)
\ee
where the integral is over $\phi_0^J$. 

The NBWF prior $P_{NB}(\phi_0^J)$ over coarse grained backgrounds\footnote{{\nf The NBWF distribution over homogeneous saddle points in \eqref{Pobs2} arises as the result of a coarse-graining over perturbations. {\kf Hence the contribution from each homogeneous saddle point in the distribution} should be interpreted as} the {\it sum} of the probabilities of an ensemble of perturbed, classical histories with widely different structures on scales associated with eternal inflation.} in \eqref{Pobs2} implies a relative weighting of landscape directions and of inflationary backgrounds within a given direction. It favors backgrounds starting at a low value of the potential followed by only a few efolds of slow-roll inflation. However, the TD weighting in \eqref{Pobs2} strongly suppresses the contribution from universes starting below the threshold of eternal inflation $\phi_{ EI}^J$. Instead it favors saddle points in regions of eternal inflation, simply because there are more possible locations of our past light cone on the resulting surfaces $\Sigma$ of constant density. 

In landscape directions where the potential has a regime of eternal inflation the low no-boundary probability of histories starting above the threshold for eternal inflation is more than compensated for by the TD weighting \cite{HHH10b,H13}. In particular in eternal inflation directions the distribution $P_{TD} P_{NB}$ over backgrounds is sharply peaked around $\phi_{EI}$. This means that the relative weighting of landscape regions of eternal inflation in \eqref{Pobs2} is specified by the potential near the eternal inflation threshold in different directions. Further, landscape directions without a regime of eternal inflation are strongly suppressed relative to directions with eternal inflation because $P_{TD}(\phi_0^J) \ll 1$ for all $\phi_0^J$ in those directions. We can therefore approximate \eqref{Pobs2} by restricting the sum over $J$ to eternal inflation directions and by selecting the background with $\phi_0^J = \phi_{EI}^J$ in each of those directions. This yields
\be
P({\cal O}|\Dobsaoi) \approx \sum_{J_{EI}^{\ }} g(J) P({\cal O}|\phi_{EI}^J) P_{NB}(\phi_{EI}^J)
\label{res}
\ee
where $J_{EI}^{\ }$ labels the directions with eternal inflation. Since the TD weighting \eqref{td} is universal it is plausible that regions of eternal inflation dominate the probabilities for observations in a large class of landscape models, including models with false vacua directions and multi-field patches.

Eq. \eqref{res} applies to all observables ${\cal O}$ that do not explicitly depend on the perturbation variables $\zeta$. Since we will be interested in the probabilities for $\Lambda$ and $Q$ below it is useful to rewrite $P_{NB}$ in terms of those variables.

We first deal with directions in which the potential takes a power law form $V(\phi) = \Lambda+\lambda \phi^n$. Here inflation can occur for $\phi \gg 1$ and ends when the slow-roll conditions break down at $\phi_e \approx n/\sqrt{2}$. In each power law direction the NBWF predicts a one-parameter family of inflationary backgrounds which can be labeled by $\phi_0 \geq \phi_{0c}$ where $\phi_0$ is the absolute value of $\phi$ at the SP of the saddle point and the lower bound $\phi_{0c} \sim {\cal O}(n)$. The number of efolds is given by $N \approx \phi_0^2/2n$. The eternal inflation threshold where $V \sim \epsilon$ is given by\footnote{We ignore factors of order one in the estimates below.}
\be
\label{exit}
\phi_\EI (\lambda, n) \approx \left(\frac{n^2}{\lambda}\right)^{\frac{1}{n+2}} ,
\ee
and hence
\be
\label{vei}
V(\phi_{EI}) \approx n^{\frac{2n}{n+2}} \lambda^{\frac{2}{n+2}}.
\ee
One of the parameters in power law potentials can be traded for the observable $Q$. For power law potentials of the form \eqref{pwlaw}
\be
\label{Qdef}
Q^2 \approx (2 n N_*)^{(n+2)/2} \frac{\lambda}{n^2},
\ee
where $N_*$ is the number of inflationary efolds between reheating and when the COBE scale left the horizon --- a number of order $50$. 
Substituting \eqref{Qdef} in \eqref{vei} gives the NBWF prior $P_{NB}(\phi_{EI}^J)$ in \eqref{res} in power law directions of eternal inflation labeled by $J_{EI}^{\ } = (\Lambda,Q,n)$,
\be
P_{NB} (\phi_{EI}^J) = \exp \left[\frac{3\pi}{\Lambda+ V(\phi_{EI}^J)}\right]  \approx \exp\left[\frac{3\pi}{\Lambda+nQ^{\alpha}/4N_*}\right]
\label{pwprior}
\ee
where
\be
\label{qdef}
\alpha \equiv \frac{4}{2+n},  \quad   0<\alpha<2 .
\ee

The analysis is similar for the plateau-like directions in our model landscape with $V(\phi) = \Lambda+V_0 (1-\phi^p/\mu)^2$.
These constitute a four parameter family of directions labeled by $J=(\Lambda, V_0, \mu, p)$. The threshold for eternal inflation occurs at 
\be
\label{exitplat}
\phi_{EI} \approx \left( \frac{\mu^2 V_0}{p^2}\right) ^{\frac{1}{2p-2}}
\ee 
and $V(\phi_{EI}) \approx V_0$. In terms of the potential parameters we have
\be
\label{Qplat}
Q^2 \approx \frac{2(p-2)N_{*}V_0}{p(\phi_{*}^p/\mu)}, \quad p \geq 3 
\ee
and
\be
Q^2 \approx V_0 e^{2N_{*}/\mu}, \quad p=2
\ee
Trading $V_0$ for $Q$ in the NBWF prior then yields
\begin{subequations}
\label{plamq}
\be
P_{NB} (\phi_{EI}^J) \approx  \exp\left[\frac{3\pi}{\Lambda + nQ^2(\phi_{*}^p/\mu)/2(p-2) N_{*}}\right], \quad p \geq 3 ,
\label{platprior1}
\ee
and
\be
P_{NB} (\phi_{EI}^J) \approx \exp\left[\frac{3\pi}{\Lambda + Q^2e^{-2N_{*}/\mu}}\right], \quad p=2 . 
\label{platprior2}
\ee
\end{subequations}

The distribution $P_{NB}(\phi_{EI}^J)$ is approximately flat over $\Lambda$ in our model landscape since $\Lambda \ll Q$ in the anthropically allowed range. By contrast it exhibits a {\it universal} exponential behavior $\exp[{\rm const}/Q^{\alpha}]$, with $\alpha =2$ for plateau-like potentials and $\alpha <2$ for power laws\footnote{An exponential prior on $Q$ was previously discussed in \cite{Garriga05}.}. This selects the plateau-like potentials with a regime of eternal inflation in our landscape. 

{\qf We proceed in the next section with a restricted two-parameter set of plateau directions \footnote{It turns out our results don't depend much on the precise values adopted for $p$ and $\mu$.} with fixed $p$ and $\mu$ but with varying $Q$ and $\Lambda$. The top-down probabilities in this simplified landscape take the form
\be
\label{general}
P({\cal O}|\Dobsaoi) \approx \sum_{\Lambda,Q} P({\cal O}|\Lambda,Q) P_{NB}(\Lambda,Q)
\ee
where $P_{NB}(\Lambda,Q) \approx \exp[c/Q^2]$ with $c$ a constant.

%%%%%777777777777777777777777777777777777777777777777777777777777777777777777777
\section{Anthropic Selection in Large Universes}
\label{ourHV}

This section applies the general result \eqref{general} for the first person, top-down probabilities predicted by the NBWF to a discussion of the correlations predicted among the observables $\cO=(\Lambda,Q,\Dell)$ in our Hubble volume. These correlations contain those examined by Weinberg \cite{Wei89} and others \cite{Banks04,Graesser04,LR05,Garriga05,Feldstein05,Tegetal06} assuming uniform priors. Here we are inevitably led to the  prior {\kf supplied by the universe's quantum state. The consequences are illustrated here by the example of the no-boundary quantum state \eqref{plamq}. }

For the observables $(\Lambda,Q,\Dell)$ equation \eqref{general} becomes
\be
\label{bayes}
P(\Lambda, Q, \Dell|\Dobsaoi)= P^{(1p)}(\Lambda, Q, \Dell) =P(\Dell|\Lambda, Q) P_{NB}(\Lambda,Q) .
\ee
{\mf Thus we have derived from the theory $\FT$ the starting point \eqref{bayes1} for the short summary of the derivation of anthropic bounds given in the Introduction. This section follows that outline but with more care and greater detail.}

Using the TAR terminology, $P(\Dell|\Lambda, Q)$ is the selection probability and $P_{NB}((\Lambda,Q)$ is the prior. Both of these have a well defined quantum mechanical origin in the fundamental theory and neither is an expression of ignorance as in TAR. {\mf Further}, the correlation calculated is more  general than usually studied in TAR. This is because $\Dell$ includes data that are not crucial for life or for our existence --- the time $t_0$ for instance. Rather these probabilities are `anthropic' in a more general sense: They refer to {\it our} data in {\it our} Hubble volume. This is discussed in a more general context in  Section \ref{testing}. 

The probabilities $P(\Dell|\Lambda, Q)$ will be proportional to the number of habitable galaxies that have formed in our Hubble volume by $t_0$. In principle this number can be calculated from the Gaussian NBWF probabilities of the quantum fluctuations $\zeta$. Here we approximate this by the expected number of galaxies for given values $(Q,\Lambda)$, which follows from the classical equations of motion. This approximation connects our analysis to classical discussions of anthropic selection in particular those in \cite{Tegetal06}. We include our data on {\tf the age of galaxy formation} $t_0$ in a similar manner by requiring that $t_0$ is greater than the classical collapse time  $t_C(\Lambda,Q)$ of a typical fluctuation. This leads to the following expression for $\SP$ characteristic of anthropic selection [cf. \eqref{fractgal}],
\be
\label{fractgal}
P(\Dell|\Lambda,Q) = P_g(\Dell) f(\Lambda, Q, t_0)\theta(t_0-t_C(\Lambda,Q)) .
\ee
Here, $\theta(x)$ is the step function,  $f(\Lambda, Q, t_0)$ is the fraction of protons in our Hubble volume in the form of galaxies by {\pf the time} $t_0$, and $P_g(\Dell)$ is the probability that any one galaxy has the specific features of ours contained in the data $\Dell$.  This is plausibly independent of $\Lambda$ and $Q$ and, like $P_g(\Dobs)$ in Section \ref{small}, will not  appear in the final probabilities. 

%%%%%%%%%%%%%%%jj
\begin{figure}[t]
\includegraphics[width=4.0in]{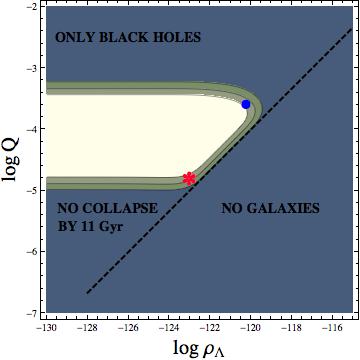} 
\caption{The  $\Lambda$-$Q$ plane. This figure, adapted from \cite{Tegetal06} (the same as Figure \ref{fraction1}) shows the various factors affecting the calculation of the selection probability $P(\Dell|\Lambda, Q)$ that is the basis for anthropic selection. The region that allows for gravitationally collapsed bodies  at some time is to the left of the diagonal dotted line assuming the presently observed matter density.  The contours trace constant values of the fraction $f(\Lambda, Q, t_0)$ of protons in the form of galaxies at {\pf a time  $t_0 \sim11$ Gyr a few billion years before the present age}, as calculated from formulae in \cite{Tegetal06}. The displayed curves correspond to $f=.4,.5,.6$. The selection probability $P(\Dell|\Lambda, Q)$ is approximately uniform in the interior of the contours shown and decreases quickly outside them. The most probable values for
$\Lambda$ and $Q$ are indicated by the  {\color{red}\bigstar}\  assuming the no-boundary prior \eqref{plamq}. These coincide with the observed values. The correlation thus supports the theory that led to it.  The most probable values for a uniform prior in $Q$ and $\Lambda$ are at the largest value of $\log \Lambda$ in the allowed range and indicated by the {\color{blue}\bigdot}.  These are not close to the observed values.}
\label{fraction}
 \end{figure}
%%%%%%%%%%%%%%%

A contour plot of $f(\Lambda, Q, t_0)$ for $t_0 \sim 11 \ {\rm Gyr}$, {\pf a few billion years before the present age}, is shown in Figure \ref{fraction}, calculated along the lines of \cite{Tegetal06}. The fraction involves the combination $\Lambda/Q^3$. This gives rise to the dotted line with slope 3 in Fig \ref{fraction}. For combinations $(\Lambda, Q)$ to the right of this line the universe expands too quickly for galaxy-sized, gravitationally bound systems to form. For $Q \geq 10^{-3}$ halos consist mostly of massive black holes, whereas for $Q \leq 10^{-5}$ galaxies will not have formed by  $t_0$ \cite{Tegmark98}. In both regimes the fraction $f$ of matter in habitable galaxies is essentially negligible.  

The fraction $f$  is also a function of time. For times much before the collapse time $t_C(\Lambda, Q)$ the fraction $f(\Lambda, Q, t) $ of protons in galaxies is very small. For times that are greater than $t_C(\Lambda, Q)$ the fraction approaches an asymptotic value. The sharp constraint on the the time of galaxy formation $t_0$ represented by the step function in \eqref{fractgal} can thus be approximately implemented\footnote{\nf Like Tegmark et al. in \cite{Tegetal06} we use the characteristic density $\rho_{vir}$ of halos as a time variable in the fraction, because this is the anthropically relevant parameter that determines whether galaxies are habitable (i.e. contain stars); a minimum $\rho_{vir}$ is required for stars to form. Halos that form later in a given history have a lower density $\rho_{vir}$. Within the anthropically allowed range for $\Lambda$ an upper bound on the collapse time $t_C$ is approximately equivalent to the requirement of a lower bound on $\rho_{vir}$.} by a sharp constraint that the fraction $f$ be above some value $f_C$ by $t_0$.

{\wf It is instructive to compare the large universe result in \eqref{bayes} and \eqref{fractgal} with the small universe result in \eqref{tar}. In the latter selection was by $D_r=(\Dobs,\Dell)$. Here it is only by $\Dell$. This a significant difference in amount of data. But it is not a great difference in selective power because the probabilities for $\Dobs$ do not depend on $\Lambda$ and $Q$. Whatever data is used {\nf to specify the observational situation in large universes} will not be available for selection because it occurs with unit probability somewhere in an infinite universe. {\nf By describing our observational situation only in terms of ultra local data $\Dobs$} we have preserved the maximum data for anthropic selection in \eqref{bayes} and \eqref{fractgal}. }

%%%%888888888888888888888888888888888888888888888888888888888888888888888888888888888
\section{Anthropic Predictions for $\Lambda$}
\label{together}

We now show that the NBWF predicts a tight correlation between $\Lambda$, $Q$ and $\Dell$ in our model.
This correlation can be expressed in terms of the joint probabilities \eqref{bayes}.

The probability for the data $\Dell$ is negligible outside the small region in the $(\Lambda,Q)$-plane where the fraction of matter in galaxies by the present age is significant. Within the selected region $\SP$ varies only slowly.  But the NBWF prior \eqref{plamq} exponentially favors the smallest allowed value of $Q$ within this region. The product \eqref{bayes} is therefore strongly peaked at $Q\sim{\cal O}(10^{-5})$ {\wf because of this exponential behavior and the sharp age cutoff in \eqref{fractgal}.} The NBWF prior has thus effectively transformed the two parameter problem into a problem for the one remaining parameter $\Lambda$. The NBWF prior in \eqref{plamq} is very close to uniform over the tiny allowed range for $\Lambda$ from $0$ to $\sim 10^{-123}$. In effect we have recovered Weinberg's original setup \cite{Wei89} in which $Q$ was fixed to its observed value and $\Lambda$ allowed to vary with a uniform distribution over $\Lambda$. 

Since we are only interested in the order of magnitude of $\Lambda$ it is natural to recast this uniform distribution in terms of $\log_{10}\Lambda$. The resulting distribution is shown in the first graph in Figure \ref{peak}. With Weinberg we can conclude that it is likely that $\Lambda$ is within one or two orders of the upper limit of the range $\Lambda \sim {\cal O}(10^{-122})$.  That is close to the observed value of $\sim 10^{-123}$. Thus we predict with significant probability a correlation between the observed values of $\Lambda$, $Q$, and the data $\Dell$ at the present age.

%%%%%%%%%%%%%%%%%%%%%%%%%%%%%%%%%%%%%%%%%%%%%%%%
 \begin{figure}[t]
\includegraphics[width=3.0in]{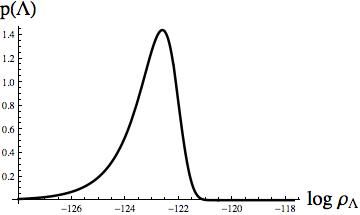} 
\includegraphics[width=3.0in]{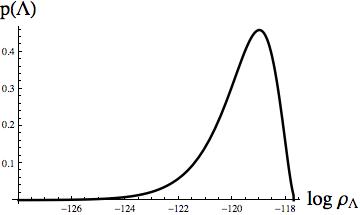}
\caption{Probability densities for $\log_{10}(\Lambda)$ . The first curve shows the distribution of $\log_{10}(\Lambda)$ resulting from marginalizing the probability \eqref{bayes} over $Q$ with the NBWF prior. The peak of the distribution is at $\Lambda \sim \cO(10^{-123})$ which in order of magnitude is  the observed value. The second curve shows the same distribution with a uniform prior replacing the NBWF. The peak is a few orders of magnitude above the observed value.}
\label{peak}
 \end{figure}
 %%%%%%%%%%%%%%%%%%%%%%%%%%%%%%%%%%%%%%%%%%%%%%%%%%%%

Were the no-boundary prior $P_{NB}(Q,\Lambda)$ replaced by the uniform prior commonly used in anthropic arguments the results would be different. As the second graph in Figure \ref{peak} shows, the largest probability in the allowed region would occur at $\Lambda \sim{  \cal O} (10^{-119})$ \cite{LR05,Tegetal06}. The correlation that emerged using the NBWF prior would not predicted. 

{\wf The NBWF prior predicts a correlation that agrees roughly with observation and the uniform prior does not. However our model is too simple to infer that this will generally be the case. The important point is that the results are different. We discuss what this means in the next section.}

%%%%%%%%%9999999999999999999999999999999999999999999999999999999 
\section{Testing Theory in Quantum Cosmology}
\label{testing}

Theories are tested, supported, and falsified by comparing their predictions with observation. What is important for this are not the bottom-up probabilities for entire classical histories that follow directly from $\FT$. Rather what is important are the top-down probabilities for local observables  conditioned on a description of the observational situation and coarse grained over what is not observed. These have been the focus of this paper.

An objective program for supporting a theory in cosmology is to search among all possible top-down correlations in {\wf our total  data $D$  on all scales (including the results of observations) for ones that are predicted by $(I,\Psi)$  with significant probability and are sensitive to the theory's details. The observables participating in the correlation must be selected to this end.  Not every correlation will test the theory\footnote{For example, correlations between a record of an observation and the observed value.}. }  But we posit no list of correlations that must have significant probability for a successful theory\footnote{The process of making predictions of future observations can be handled by augmenting $D$ by their possible results.}.  We search through all. We reject for example the idea that the fundamental laws of physics $(I,\Psi)$ must  be such as to make human observers probable or typical of anything \cite{HS07}.

The successful prediction of a correlation between $\Lambda$, $Q$, and $\Dell$ discussed in this paper is an example of one that supports the NBWF {\kf as a theory of the universe's quantum state. {\pf This result, as well as related studies of correlations involving CMB fluctuation observables \cite{H13}, suggest different theories of the quantum state can to some extent be observationally distinguished and thereby tested. On the other hand we have no evidence that the NBWF is the unique state exhibiting this correlation\footnote{Holography provides some theoretical support for a unique wave function of the universe \cite{HM04}.}.} Other theories of the state may do just as well, or be supported by successful predictions of other correlations.} 

 In this section we argue why this particular correlation {\kf predicted by the NBWF} was fruitful by discussing two examples of what would happen if some of these observables had been left out.

\vskip.2in
1. {\it Leaving out the Observer }
\vskip.2in

The first example concerns $\Dobs$ describing us and our observational situation} as physical systems within the universe. It is striking that the details of this data turned out to be unimportant for exhibiting the correlation. But it was nevertheless crucial for the result to take in account the observational situation. A common intuition is that small systems like us have a negligible influence on the evolution of the universe and can be safely ignored in cosmology. {\nf That is true for the evolution of each classical history individually. It is also true for the bottom-up probabilities for the ensemble of classical histories that follow from $\FT$.  But it is not the case for the (top-down) probabilities for the results of our observations \cite{HH06}. The latter are necessarily conditioned on a description of the observational situation. As discussed in Section \ref{pred} this generally has a large effect on the results.} We have seen that the top-down probabilities conditioned on at least one instance of $\Dobs$ favor the large universes generated by eternal inflation because they have more places for $\Dobs$ to occur. The systems described by $\Dobs$ are indeed small but  they have a big effect on the predictions of the theory for observation in a large universe precisely because the probabilities for them are so small. They are crucial to selection.

\vskip.2in
2. {\it Leaving Out the Time $t_0$}
\vskip.2in

An example related to the discussion in Section \ref{ourHV} concerns the time $t_0$.  Suppose we had calculated the joint probability describing a correlation between $\Lambda$, $Q$, and {\nf a more restricted set of} local data $\Dloct$ that do not include a specification of the age, for instance by not including the Hubble constant. After all, $t_0$  is not a traditional anthropic variable because it need not take a particular value for life to exist. But quantum cosmology is not restricted to a set of traditional anthropic variables. Rather it predicts probability distributions for any set of cosmological alternatives. Life is just another kind of physical system within the universe.

To examine whether the NBWF predicts a correlation between the observed values of $\Lambda$ and $Q$ and the observation that $\Dloct$ exists {\it some time} one must evaluate the joint probability \eqref{bayes}, where $P(\Dloct|\Lambda, Q)$ is again of the form \eqref{fractgal} but now with the fraction $f$ replaced by its asymptotic value and without the step function in time. Figure \ref{noage} shows the region in the $\Lambda-Q$ plane where the probabilities $P(\Dloct|\Lambda, Q)$ are significant. The lower bound on $Q$ is no longer set by the time $t_0$ as in Figure \ref{fraction}, but by the requirement that there be sufficient radiative cooling of collapsed halos for stars to form \cite{Tegetal06}. 

%%%%%%%%%%%%%%%
\begin{figure}[t]
\includegraphics[width=4.0in]{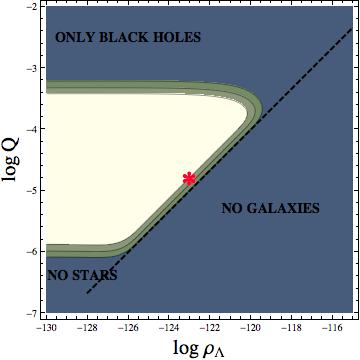} 
\caption{The  $\Lambda$-$Q$ plane when observations of the age are not taken into account. If galaxies are not required to form by  the present age then the anthropically allowed region in which galaxies of stars can form some time extends to smaller values of $Q$ than it does in Figure \ref{fraction} with an age constraint. The figure shows the allowed region as discussed in the text. The \mbox{\Large $* $} locates the observed values of $\Lambda$ and $Q$. 
}
\label{noage}
 \end{figure}
%%%%%%%%%%%%%%%

The NBWF prior \eqref{plamq} favors the smallest $Q$ in the region selected by $P(\Dloct|\Lambda, Q)$ which is $Q\sim 10^{-6}$. In those universes galaxies typically form at a time later than $t_0$ from the collapse of perturbations which have an initial amplitude $Q$ smaller than the observed value. The NBWF prior assigns much lower probability to a correlation between the observed values of $\Lambda$, $Q$ and the existence of $\Dloct$ some time, than it does to the correlation between $\Lambda$, $Q$ and $\Dell$ that includes a specification of the {\tf present age and the time of galaxy formation  $t_0$}. Does this mean that the theory has failed in its predictive power? No, because in quantum cosmology we expect only some correlations to be predicted with significant probabilities. We have to search to find these, and the comparison of these two correlations is an example of this search. 

As can be seen from Figure \ref{noage} the NBWF prior assigns a significant probability  to a correlation between $\Dloct$ existing some time and the values $Q\sim 10^{-6}$, $\Lambda \sim 10^{-127}$ both not what we observe. Does this falsify the theory? No, because a piece of known data that significantly affects the results has been neglected in the calculation --- the time $t_0$. It would be like an experiment determining the value of some constant failing to correct for some known effect that significantly affects the result. Data that we have should not be left out unless it can be demonstrated that they do not affect the  result. Indeed, if we had neglected the formation of stars in addition to the age,  $Q$ would have been pushed down to $\co (10^{-10})$.

%%%%%10101010101010101010101010
\section{Summary --- A Series of Selections} 
\label{summary}

In this paper we used a toy model of a final theory to calculate the probability for a correlation between three pieces of our data about the universe acquired by both large and small scale observations. The three pieces were the cosmological constant $\Lambda$, the parameter $Q$ summarizing the size of fluctuations in the CMB, and data $\Dell$ describing  our nearby universe of galaxies including data that determine our present time from reheating. The toy model assumed dynamics described by a multi-dimensional scalar field potential defining a landscape consisting of different minima that can be approached from different one-dimensional directions in field space. The semiclassical NBWF was assumed as the model of the universe's quantum state. 

{\uf The quantum state predicts {\qf a multiverse that includes} an ensemble of classical histories of the universe along with their quantum probabilities. The probability that we will observe a particular value of $\Lambda$ and $Q$ involves the probability that our past history rolled down to a minimum characterized by $\Lambda$ in a direction characterized by $Q$. This paper's calculation of this probability} can be conveniently summarized as a series of {\qf selections of histories in the multiverse}  --- either made by us in deciding what to calculate, or by probabilities that favor one range of parameters over another. 

\begin{itemize}

\item{\it Selection for Classical Histories:}  We assume our data are part of a quasiclassical realm of the wave function of the universe described by an ensemble of alternative histories with high probabilities from the theory for classical deterministic evolution\footnote{In particular this means that we assume that we are not Boltzmann brains. See \cite{HS07} for more on how to do this.}. Patches of the landscape that do not contribute to the classical ensemble of universes are therefore excluded by assumption. In the context of the NBWF the existence of classical space-time selects those patches of the landscape where the slow-roll conditions for inflation hold.

\item{\it Selection for Eternal Inflation:} Probabilities for observation are necessarily conditioned on the existence of at least one instance of some part of our data  {\qf somewhere in the universe}. As discussed in Section \ref{pred} this selects the very large histories with a regime of eternal inflation. 

\item{\it Selection for the Lowest Exit  from Eternal Inflation:} The NBWF prior favors eternally inflating universes emerging in landscape regions where the threshold for eternal inflation $V=\epsilon$ lies at a low value of the potential. The structure of the landscape beyond the eternal inflation threshold is irrelevant for our observations in the NBWF.

\item{\it Anthropic Selection:} We won't observe parameters characterizing histories or {\qf regions of histories} for which there is a negligible probability that we exist. {\qf Joint probabilities for local observables that include us therefore select histories and landscape regions where the probability that we exist is significant.}

\item{\it Non-anthropic Selection:}  As always in a top-down analysis, the selected histories of the universe depend on the precise question asked \cite{HH06}. When a correlation involves non-anthropic observables as well as anthropic ones the probabilities for the whole set acts as a selection mechanism on the landscape. The present calculation is an example with the {\tf the galaxy formation time}  $t_0$ being non-anthropic. 
 
\end{itemize}

The result of the calculation was that Weinberg's argument for $\Lambda$ was effectively restored even though $Q$ varies in the landscape. The model yielded specific values for $\Lambda$ and $Q$ that were in order of magnitude consistent with those observed. However, this crude agreement with observation is not our main result. The model 
is too simple for that. Rather we have two main conceptual results. First, anthropic selection is naturally and inevitably a part of quantum cosmology if we are interested in predictions for observations by physical systems within the universe. Second, the probabilities for anthropic selection are not determined by some assumed prior reflecting ignorance of the final theory, but rather by the quantum state of the universe that is part of the specification of the final theory. 
 
Anthropic selection is thus not some special process  that is separate from the final theory whose employment or not is to be debated. In our analysis it is but one of a series of selections described above. It is an inevitable part of the calculation of quantum mechanical probabilities for the predictions of our observations when the fact that observers are physical systems within the universe is taken in account.

\vskip .2in

\noindent{\bf Acknowledgments:} We thank Martin Rees and Mark Srednicki for stimulating discussions. TH thanks the KITP and the Physics Department at UCSB for their hospitality.  JH thanks Marc Henneaux and the International Solvay Institutes for their hospitality. The work of JH was supported in part by the US NSF grant PHY12-05500. The work of TH was supported in part by the US NSF Grant PHY08-55415, {\qf by Joe Alibrandi} and by the National Science Foundation of Belgium under the FWO-Odysseus program.


\begin{thebibliography}{99}

\bibitem{Barrow86} J. Barrow, F. Tipler, \ttle{\it The Anthropic Cosmological Principle,} Oxford University Press, Oxford (1986)

\bibitem{Wei89} S.~Weinberg, \ttle{\it Anthropic Bound on the Cosmological Constant,} Phys.~Rev.~Lett., {\bf  59}, 2607, (1987)

\bibitem{Banks04} T. Banks, M. Dine, E. Gorbatov, \ttle{\it Is there a string theory landscape?,} JHEP {\bf 0408}, 058 (2004), arXiv:hep-th/0309170

\bibitem{Graesser04} M. Graesser, S. D. H. Hsu, A. Jenkins, M. B. Wise, \ttle{\it Anthropic Distribution for Cosmological Constant and Primordial Density Perturbations,} Phys. Lett. B {\bf 600}, 15 (2004), arXiv:hep-th/0407174

\bibitem{LR05}  M.~Livio and M.J.~Rees, \ttle{\it Anthropic Reasoning,} Science, {\bf 309}, 1022 (2005).

\bibitem{Garriga05} J. Garriga, A. Vilenkin, \ttle{\it Anthropic Prediction for $\Lambda$ and the Q catastrophe,} Prog. Theor. Phys. Suppl. {\bf 163}, 245 (2006), arXiv:hep-th/0508005

\bibitem{Feldstein05} B. Feldstein, L. J. Hall, T. Watari, \ttle{\it Density Perturbations and the Cosmological Constant from Inflationary Landscapes,} Phys. Rev. D {\bf 72}, 123506 (2005), arXiv:hep-th/0506235

\bibitem{Tegetal06} M.~Tegmark, A.~Aguirre, M.J.~Rees, and F. Wilczek,  \ttle{\it Dimensionless constants, cosmology, and other dark matters,} Phys. Rev. D {\bf 73}, 023505 (2006). 

\bibitem{Schellekens13} A. N. Schellekens, \ttle{\it Life at the Interface of Particle Physics and String Theory,} arXiv:1306.5083

\bibitem{HH06} S.W. Hawking and T. Hertog, \ttle{\it Populating the Landscape: A Top Down Approach,} Phys. Rev. D {\bf 73}, 123527 (2006), arXiv:hep-th/0602091

\bibitem{Har05} J. B. Hartle, \ttle{\it Anthropic Reasoning and Quantum Cosmology,} AIP Conf. proc. {\bf 743}, 298 (2005), arXiv:gr-qc/0406104

\bibitem{HHH10b} J. B.~Hartle, S. W. Hawking, T. Hertog, \ttle{\it Local Observation in Eternal Inflation,} Phys. Rev. Lett. {\bf 106}, 141302 (2011), arXiv:1009.2525

\bibitem{H13} T.~Hertog, \ttle{\it Predicting a Prior for Planck}, arXiv:1305.6135

\bibitem{HH83} J. B.~Hartle and S. W.~Hawking, \ttle{The Wave Function of the Universe,} Phys. Rev. D {\bf 28}, 2960-2975 (1983).

\bibitem{HHH08} J. B.~Hartle, S. W. Hawking, T. Hertog, \ttle{\it Classical universes of the no-boundary quantum state,} Phys. Rev. D {\bf  77}, 123537 (2008), arXiv:0803.1663

\bibitem{HHH08b} J. B.~Hartle, S. W. Hawking, T. Hertog, \ttle{\it The no-boundary measure of the universe,} Phys. Rev. Lett. {\bf 100}, 202301 (2008), arXiv:0711:4630.

\bibitem{HH09}  J.B.~Hartle and T. Hertog, \ttle{\sl Replication regulates volume weighting in quantum cosmology,} Phys. Rev. D {\bf 80}, 063531 (2009),  arXiv:0906.0042. 

\bibitem{Page97} D. Page, \ttle{\it Space for both No-Boundary and Tunneling Quantum States of the Universe,} Phys. Rev. D {\bf 56}, 2065 (1997). 

\bibitem{Perlmutter99} Supernova Cosmology Project (S. Perlmutter {\it et al.}), \ttle{\it Measurements of Omega and Lambda from 42 high redshift supernovae,} ApJ {\bf 517}, 565 (1999(, arXiv:astro-ph/9812133 

\bibitem{Planck13} P. A. R. Ade et al., \ttle{\it Planck 2013 results XXII: Constraints on Inflation,} arXiv:1303.5082

\bibitem{Bousso00} R. Bousso, J. Polchinski, \ttle{\it Quantization of four-form fluxes and dynamical neutralization of the cosmological constant,} JHEP {\bf 06}, 006 (2000), arXiv:hep-th/0004134

\bibitem{Denef04} F. Denef, M. Douglas, \ttle{\it Distributions of flux vacua,} JHEP {\bf 05}, 072 (2004), arXiv:hep-th/0404116.

\bibitem{HH90} J.~Halliwell and J.B.~Hartle, {\it Integration Contours for the No-Boundary Wave Function of the  Universe}.  {\sl Phys. Rev. D}, {\bf 41},1815  (1990).

\bibitem{HL90} J.~Halliwell and J~Louko, {\it Steepest-Descent Contours in the Path-Integral Approach to Quantum Cosmology. III A General Method with Application to Anisotropic Minisuperspace Models.}, {\sl Phys. Rev. D} {\bf 42}, 3997 (1990).

\bibitem{HM04}
G. T. Horowitz, J. M. Maldacena, \ttle{\it  The black hole final state}, JHEP {\bf 0402}, 008 (2004), hep-th/0310281

\bibitem{HH11} T. Hertog and J.B. Hartle,  \ttle{\it Holographic No-Boundary Measure}, JHEP {\bf 1205}, 095 (2012), arXiv:1111.6090

\bibitem{GH07} M. Gell-Mann and J.B~Hartle, {\it Quasiclassical Coarse Graining and Thermodynamic Entropy}, {\sl Phys. Rev. A}, {\bf 76}, 022104 (2007), arXiv:quant-ph/0609190.

\bibitem{HHH10a} J. B.~Hartle, S. W. Hawking, T. Hertog, \ttle{\it The No-Boundary Measure in the Regime of Eternal Inflation,} Phys. Rev. D {\bf 82}, 063510 (2010), arXiv:1001.0262

\bibitem{Halliwell85} J. J. Halliwell, S. W. Hawking, \ttle{\it Origin of Structure in the Universe,} Phys. Rev. D {\bf 31}, 1777 (1985).

\bibitem{Gel94} M.~Gell-Mann, {\sl The Quark and the Jaguar}, W.~Freeman
San Francisco (1994).

\bibitem{HS09} M.~Srednicki and J.B.~Hartle, {\it Science in a Very Large Universe}, {\sl Phys. Rev. D}, {\bf 81} 123524 (2010),  arXiv:0906.0042, {\it The Xerographic Distribution: Scientific Reasoning in a Large Universe}, arXiv:1004.3816. 

\bibitem{Tegmark98} M. Tegmark, M. J. Rees, \ttle{\it Why is the CMB fluctuation level $10^{-5}$?,} ApJ {\bf 499}, 555 (1998), arXiv:astro-ph/9709058

\bibitem{HS07} J.B. Hartle and M. Srednicki, \ttle{\it Are We Typical?,} Phys. Rev. D {\bf 75}, 123523 (2007), arXiv:0704.2630. 

\end{thebibliography}
\end{document}